\documentclass[twocolumn,showpacs,preprintnumbers,amsmath,amssymb]{revtex4-1}

\usepackage[english]{babel}
\usepackage[T1]{fontenc}
\usepackage{bbm}
\usepackage{epsfig}
\usepackage{psfrag}
\usepackage{color}
\usepackage{amsmath}
\usepackage{amsfonts}
\usepackage{amssymb}
\usepackage{graphicx}%
\newcommand{\be}{\begin{equation}}
\newcommand{\ee}{\end{equation}}
\newcommand{\bea}{\begin{eqnarray}}
\newcommand{\eea}{\end{eqnarray}}

\newcommand{\Q}{{\bf a}}

\begin{document}

\title{Exact self-accelerating cosmologies  in the ghost-free 
bigravity and massive gravity
}

\author{Mikhail~S.~Volkov}

\affiliation{
Laboratoire de Math\'{e}matiques et Physique Th\'{e}orique CNRS-UMR 7350, 
Universit\'{e} de Tours, Parc de Grandmont, 37200 Tours, FRANCE}

\email{volkov@lmpt.univ-tours.fr}

\begin{abstract}

Within the theory of the ghost-free bigravity, 
we present the most general cosmological solution 
for which the physical metric is homogeneous and isotropic, 
while the second metric is inhomogeneous. 
The solution includes a matter source and exists for any values 
of the theory parameters. 
The physical metric describes a universe with
the late time acceleration due to the effective cosmological term
mimicked by the graviton mass.
When perturbed, this universe should rest approximately homogeneous and isotropic
in space regions small compared to the graviton Compton length. 
In the limit where  the massless graviton 
decouples, the solution fulfills the equations of the ghost-free 
massive gravity.

\end{abstract}

\maketitle

Considering theories with massive gravitons \cite{Pauli} 
is motivated 
by the observation of the current acceleration of our universe \cite{Reiss}, 
since the graviton mass can induce an effective cosmological term. 
Although such theories can exhibit unphysical features, as for example 
the Boulware-Deser ghost \cite{BD}, the recent discovery of the special 
massive gravity \cite{RGT} and its bigravity generalization \cite{HR1} 
which are ghost-free \cite{noghost},
suggests that such theories can indeed be good candidates 
for interpreting the observational data. This motivates studying cosmological
solutions with massive gravitons.

The first self-accelerating cosmologies  in the ghost-free massive gravity 
were obtained without matter and describe the pure de Sitter universe \cite{Koyama}. 
The matter source was then included, but only for special values of the theory 
parameters \cite{cosm}. 
For these solutions the physical metrics is of the Freedman-Robertson-Walker (FRW) type, 
but the fiducial metric is inhomogeneous. 
This means that perturbations around the background 
can give rise to inhomogeneity, although this effect should be suppressed
by the smallness of the graviton mass \cite{gaba}. One more similar solutions was 
found in Ref.\cite{gaba}, where it was argued that solutions of this type
should be generic for massive gravity.
The theory also admits solutions for which both metrics are FRW,
but these show a nonlinear instability and seem to be unphysical \cite{Muk}.

The generalizations of results of \cite{Koyama},\cite{cosm}
within the ghost-free bigravity were obtained in  \cite{cosm1},\cite{Volkov}.
For these solutions both metrics are dynamical but are not simultaneously diagonal,
and the second metric does not share the translational symmetries 
of the first one. 
In the bigravity too, 
only special solutions of this type are known -- 
either without matter or for constrained values of the theory parameters.

In what follows, we present the most general cosmological solution 
for which the physical metric is FRW
but the second metric is inhomogeneous. 
We construct this solution within the ghost-free bigravity of \cite{HR1},
but it describes the massive gravity case as well, since its  
second metric becomes flat in the limit where the massless graviton decouples.    
The solution includes the matter source, it exists for all values of the 
theory parameters, and its physical metric describes a FRW universe 
that can be spatially flat, open or closed, and which shows 
the late-time acceleration due to the effective cosmological term mimicked
by the graviton mass.

\noindent
{\bf The ghost-free bigravity--}
The generic bigravity theory \cite{ISS}
is defined on  a spacetime manifold 
equipped with two metrics $g_{\mu\nu}$ and  $f_{\mu\nu}$,
whose kinetic terms are chosen to be of the 
standard Einstein-Hilbert form.
To describe the ghost-free bigravity \cite{HR1}, it is convenient to 
use the tetrad formalism \cite{tetrad}, in which the inverse of $g_{\mu\nu}$
and the $f_{\mu\nu}$ are parameterized as 
\be
g^{\mu\nu}=\eta^{AB}e_A^\mu e_B^\nu,~~~~~~~
f_{\mu\nu}=\eta_{AB}\omega^A_\mu \omega^B_\nu,
\ee
with $\eta_{AB}={\rm diag}[1,-1,-1,-1]$.
The action is 
\bea
S&=&-\frac{1}{16\pi G}\,\int R\sqrt{-g}\,d^4x
-\frac{1}{16\pi {\cal G}}\,\int {\cal R}\,\sqrt{-f}d^4x  \notag \\
&+&S_{\rm int }
+S_{\rm m}\,,
\eea
where $R$ and ${\cal R}$ are the Ricci scalars for $g_{\mu\nu}$ and $f_{\mu\nu}$, 
respectively, $G$ and ${\cal G}$ are the corresponding gravitational couplings.
$S_{\rm m}$ describes the ordinary matter, 
which is supposed  to directly interact only with 
$g_{\mu\nu}$. 
The interaction between the two metrics is parameterized as 
\be                                \label{int}
S_{\rm int}=\frac{\sigma}{8\pi G}\int {\cal L}_{\rm int}\sqrt{-g}\, d^4x\,, 
\ee
where 
\bea                                           \label{lagr}
\mathcal{L}_{\rm int}&=&\frac{1}{2}((K^\mu_\mu)^2-K_{\mu}^{\nu}K_{\nu}^{\mu})
+\frac{c_{3}}{3!}%
\,\epsilon_{\mu\nu\rho\sigma}
\epsilon^{\alpha\beta\gamma\sigma}K_{\alpha}^{\mu}
K_{\beta}^{\nu}K_{\gamma}^{\rho}   \notag \\
&+&\frac{c_{4}}%
{4!}\,
\epsilon_{\mu\nu\rho\sigma}
\epsilon^{\alpha\beta\gamma\delta}K_{\alpha}^{\mu}
K_{\beta}^{\nu}K_{\gamma}^{\rho}K_{\delta}^{\sigma}\,,
\eea
$c_3,c_4$ are parameters, 
and
$K^\mu_\nu=\delta^\mu_\nu-\gamma^\mu_{~\nu}$ with 
\be                                  \label{gam}
\gamma^\mu_{~\nu}=e^\mu_A\omega^A_\nu\,. 
\ee
These expressions define the theory with two gravitons, 
one of which is massless and
another one is massive, with the mass
$m^2=\sigma\left(1+{\cal G}/{G}\right)$. 
One can introduce the angle $\eta$ such that the parameters $\sigma,{\cal G}$
are expressed as
$\sigma=m^2\cos^2\eta$ and 
${\cal G}=G\tan^2\eta$.  
Instead of $c_3,c_4$ one often uses in the literature 
$\alpha_3=-c_3/3$ and $\alpha_4=-c_4/12$.

Varying the action with respect to $e^\mu_A$ and $\omega^A_\nu$
gives the  field equations  \cite{cosm1}
\begin{align}
G^\rho_\lambda&=m^2\cos^2\eta\, T^\rho_\lambda
+8\pi G T^{{\rm (m)}\,\rho}_{~~~~\lambda} \,, \label{e1} \\
{\cal G}^\rho_\lambda&=
m^2\sin^2\eta\, {\cal T}^\rho_\lambda    \,.        \label{e2}
\end{align}
Here $G^\rho_\lambda$ and ${\cal G}^\rho_\lambda$ are the Einstein tensors 
for $g_{\mu\nu}$ and $f_{\mu\nu}$, 
respectively, while 
\be                                   \label{TTT0}
T^\rho_\lambda=\tau^\rho_\lambda
-\delta^\rho_\lambda\,{\cal L}_{\rm int}\,,~~~~~~~
{\cal T}^\rho_\lambda=-\frac{\sqrt{-g}}{\sqrt{-f}}\, 
\tau^\rho_\lambda\,,
\ee
with 
\bea					\label{tau}
\tau^\rho_\lambda&=&(\gamma^\sigma_\sigma-3)\gamma^\rho_\lambda
-
\gamma^\rho_\sigma\gamma^\sigma_\lambda 
-\frac{c_{3}}{2}%
\,\epsilon_{\lambda\mu\nu\sigma}
\epsilon^{\alpha\beta\gamma\sigma}\gamma_{\alpha}^{\rho}
K_{\beta}^{\mu}K_{\gamma}^{\nu} \notag \\
&-&\frac{c_{4}}{6}\,
\epsilon_{\lambda\mu\nu\sigma}
\epsilon^{\alpha\beta\gamma\delta}
\gamma_{\alpha}^{\rho}
K_{\beta}^{\mu}K_{\gamma}^{\nu}K_{\delta}^{\sigma}\,.  
\eea
These equations should be supplemented by the conservation condition
for the matter energy-momentum tensor, 
$
\stackrel{(g)}{\nabla}_\rho T^{{\rm (m)}\rho}_{~~~~\lambda}=0\,,
$
where $\stackrel{(g)}{\nabla}_\rho$ is the covariant derivative with respect to  
$g_{\mu\nu}$.

The field equations imply that 
$T^{\mu\lambda}=g^{\mu\rho}T_\rho^\lambda$ is symmetric,
hence so is $\gamma^{\mu\lambda}=g^{\mu\rho}\gamma_{~\rho}^\lambda$, 
which means that 
\be                             \label{const}
e_A^\mu\omega_{B\mu}=e_B^\mu\omega_{A\mu}\,
\ee
where $\omega_{A\mu}=\eta_{AB}\omega^B_\mu$. 
The latter property implies that 
$
\gamma^\mu_{~\sigma}\gamma^\sigma_{~\nu}=g^{\mu\sigma}f_{\sigma\nu},
$
therefore $\gamma^\mu_{~\nu}=\sqrt{g^{\mu\sigma}f_{\sigma\nu}}$,
which corresponds to the original theory parameterization 
\cite{RGT},\cite{HR1}.

If ${\cal G}\to 0$ then the massless graviton decouples and, if only 
$f_{\mu\nu}$ becomes  flat in this limit, the theory reduces to the massive gravity 
of \cite{RGT}.

\noindent 
{\bf Spherical symmetry--}
Introducing the spherical coordinates 
$x^\mu=(t,r,\vartheta,\varphi)$, the 
most general expression for the two tetrads 
subject to the condition \eqref{const} 
is \cite{cosm1}
\bea               
e_0&=&\frac{1}{Q}\,\frac{\partial}{\partial t },~
e_1=\frac{1}{N}\,\frac{\partial}{\partial r },~
e_2=\frac{1}{R}\,\frac{\partial}{\partial \theta },~
e_3=\frac{1}{R\sin\vartheta}\,\frac{\partial}{\partial \varphi },~\notag \\
\omega^0&=&aQ\,dt+cN\,dr,~~~~\omega^1=-{cQ}\,dt+bN\,dr, \notag \\
\omega^2&=&uR\,d\vartheta,
~~~~\omega^3=uR\sin\vartheta\, d\varphi\,,                    \label{tetrad}
\eea 
where $Q,N,R,a,b,c,u$ are functions of $t,r$. 
It is straightforward to compute $\gamma^\mu_{~\nu}$ in \eqref{gam}
and obtain the following non-zero components of  
$\tau^\mu_\nu$ in \eqref{tau}: 
\bea
\tau^0_0&=&ab+2au-3a+c^2+c_4(u-1)^2(a-ab-c^2) \notag \\
&+&c_3(u-1)(au+2ab-3a+2c^2), \notag \\
\tau^\vartheta_\vartheta&=&
u(u+a+b-3)+c_4u(1-u)[c^2+(a-1)(b-1)] \notag \\
&+&c_3u[(a+b-2)u+c^2+ab-2a-2b+3] ,
\eea
with $\tau^r_r$ obtained from $\tau^0_0$ via $a\leftrightarrow b$, 
also 
$\tau^\varphi_\varphi=\tau^\vartheta_\vartheta$, and 
\be                              \label{t01}
\tau^0_r=\frac{cN}{Q}[(c_3+c_4)u^2+2(1-2c_3-c_4)u+3c_3+c_4-3].
\ee
The interaction Lagrangian \eqref{lagr} reduces to
\bea                      \label{LLL}
{\cal L}_{\rm int}&=&u(u+2a+2b-6)+c^2+ab-3a-3b+6 \notag \\
&+&c_3(u-1)[(a+b-2)u+2c^2+2ab-3a-3b+4] \notag \\
&-&c_4(u-1)^2(c^2+ab-a-b+1).
\eea
Noting that $\sqrt{-f}/\sqrt{-g}=|e^\mu_A||\omega^A_\nu|=(ab+c^2)u^2$,
it is straightforward to evaluate the two energy-momentum tensors in 
\eqref{TTT0}. 

\noindent
{\bf Homogeneity and isotropy--}
Let us assume the metric $g_{\mu\nu}$ and the distribution of the ordinary matter
to be homogeneous and isotropic. This can be achieved by setting
$Q=N=\Q(t)$ and $R=\Q(t)f_{\rm k}(r)$ with 
$f_{\rm k}(r)=\{r,\sin(r),\sinh(r)\}$ for ${\rm k}=0,1,-1$, respectively.
We choose the matter to be of the perfect fluid type with 
$
8\pi GT^{{\rm (m)}\rho}_{~~~~\lambda}={\rm diag }[\rho(t),-P(t),-P(t),-P(t)]. 
$

Since the Einstein tensor for $g_{\mu\nu}$ is diagonal, 
so should be the
energy-momentum tensor $T^\mu_\nu$ on the right in \eqref{e1}, therefore 
one should have $T^0_r= 0$,
which requires that $\tau^0_r= 0$.
Now, $\tau^0_r$ in \eqref{t01} will vanish if either 
$c=0$, or the expression between
the brackets vanishes.  We shall be considering below the case where
$c\neq 0$ and the metrics are not simultaneously diagonal, since  
the $c=0$ case  has already been studied in detail \cite{cosm1}.

If $c\neq 0$, then $\tau^0_r$ in \eqref{t01} will vanish if 
\be                                        \label{u}
u=\frac{1}{c_3+c_4}\left( 
2c_3+c_4-1\pm\sqrt{c_3^2-c_3+c_4+1}
\right).
\ee
Inserting this into the above formulas, we find that the 
energy-momentum tensors in \eqref{TTT0} become diagonal, with 
{constant} 
$00$ and $rr$
components,
\bea
T^0_0&=&T^r_r=(u-1)(c_3u-u-c_3+3)\equiv \lambda,\notag \\
{\cal T}^0_0&=&{\cal T}^r_r=\frac{1-u}{u^2}(c_3u-c_3+2)\equiv {\tilde \lambda}.
\eea
The Bianchi identities for Eqs.\eqref{e1} then imply the conservation 
condition $
\stackrel{(g)}{\nabla}_\rho T^{\rho}_{\lambda}=0\,,
$
whose the only non-trivial component is
(we denote $\dot{}\equiv \partial_t$ and $^\prime\equiv\partial_r$)
\be                         \label{cons0}
\stackrel{(g)}{\nabla}_\mu T^{\mu}_{0}=2\frac{\dot{\Q}}{\Q}
\left(T^0_0-T^\vartheta_\vartheta\right)=0.
\ee
It is worth noting that a similar condition for ${\cal T}^\mu_\nu$ 
follows identically, due to the diff.-invariance 
of $S_{\rm int}$, 
so that there is no need to impose it separately.

Now, using the above formulas one finds 
\be
T^0_0-T^\vartheta_\vartheta=\frac{c_3u-u-c_3+2}{u-1}\,
[(u-a)(u-b)+c^2].  \label{cons}
\ee
In view of \eqref{cons0} this should vanish, so that  
either the first or the second factor on the right should be zero. 
The former case was considered in \cite{cosm}, \cite{cosm1}
(see also \cite{Kobayashi}). 
However, the condition $c_3u-u-c_3+2=0$ constraints the possible values 
of the parameters $c_3,c_4$, so that the solutions obtained in this way are not general.  
We therefore abandon this condition in what follows and require instead that 
\be                       \label{uuu}
(u-a)(u-b)+c^2=0. 
\ee
In view of this, one has
$T^0_0=T^\vartheta_\vartheta$ and ${\cal T}^0_0={\cal T}^\vartheta_\vartheta$, 
which implies  that both energy-momentum tensors are 
proportional to the unit tensor,
$T^\mu_\nu=\lambda\delta^\mu_\nu$ and 
${\cal T}^\mu_\nu=\tilde{\lambda}\delta^\mu_\nu$. 
The field equations \eqref{e1},\eqref{e2} then 
reduce to 
\begin{align}
G^\rho_\lambda&=\Lambda \delta^\rho_\lambda
+8\pi G T^{{\rm (m)}\,\rho}_{~~~~\lambda} \,, \label{ee1} \\
{\cal G}^\rho_\lambda&=
\tilde{\Lambda} \delta^\rho_\lambda    \,,       \label{ee2}
\end{align}
with 
$\Lambda=m^2\cos^2\eta\,\lambda$ and 
 $\tilde{\Lambda}=m^2\sin^2\eta\,\tilde{\lambda}$. 
As a result, 
the two metrics actually decouple one from another,
and the graviton mass gives rise to a cosmological term separately for each metric. 
However, one has to remember that solutions
of \eqref{ee1}, \eqref{ee2} 
should in addition fulfill the 
consistency condition \eqref{uuu}.

\noindent
{\bf Solution for $g_{\mu\nu}$--}
Eqs.\eqref{ee1} for $g_{\mu\nu}$ comprise a closed system, 
with no additional conditions imposed, 
so that we can solve them. The metric is  
\be                                \label{ggg}
ds_g^2=\Q^2(t)(dt^2-dr^2-
f_{\rm k}^2(r) d\Omega^2),
\ee
and the equations reduce to  
\be
3\frac{\dot{\Q}^2+{\rm k}\Q^2}{\Q^4}=\Lambda+\rho,
\ee
where $\rho$ is determined by the matter conservation condition, 
$\dot{\rho}+3({\dot{\Q}}/{\Q})(\rho+P)=0$. 
This describes a universe filled with the ordinary matter and 
containing the 
cosmological term mimicked by the graviton mass. 
At early times the matter density $\rho$ dominates, but 
at late times the cosmological term 
wins, which leads  to the self-acceleration.

\noindent
{\bf Solution for $f_{\mu\nu}$--}
Eqs.\eqref{ee2} should 
determine the metric 
\be                       \label{f}
ds_f^2=\Q^2\,(a\,dt+c\, dr)^2-\Q^2\,(b\, dr-c\,dt)^2
-U^2d\Omega^2.
\ee
Here $a,b,c$ are free
functions of $t,r$, but $U=uR(t,r)$ is already 
fixed by the above considerations. In addition, $a,b,c$ should satisfy 
the constraint \eqref{uuu}. One could therefore wonder if 
the system is not overdetermined and 
the freedom is 
enough to fulfill  all conditions. 

 To see that the latter
is indeed the case, we notice that the function $U$ can be considered  as
the new radial coordinate. The temporal coordinate can also be changed,
in such a way that in new coordinates $T,U$ the metric becomes 
diagonal. The source term in \eqref{ee2} is invariant under
reparameterizations. Therefore, the problem reduces to solving 
the Einstein equations with the 
cosmological  constant $\tilde{\Lambda}$ to find a diagonal metric 
parameterized by the Schwarzschild coordinate $U$. 
  The solution is the (anti)de Sitter metric 
\be                              \label{AdS}
df^2=\Delta^2\, dT^2-\frac{dU^2}{\Delta^2}
-U^2d\Omega^2\,,
\ee
where  $\Delta^2=1-\tilde{\Lambda}U^2/3$. There remains 
to establish the correspondence between the $T,U$ and $t,r$ coordinates
and to fulfill the constraint \eqref{uuu}.

Let us introduce 1-forms
\bea
\theta^0={\Delta}dT,~~~\theta^1=\frac{dU}{{\Delta}},
~~~\theta^2=Ud\vartheta,
~~~\theta^3=U\sin\vartheta d\varphi\,,   \notag                 \label{tetrad1}
\eea 
such that 
$f_{\mu\nu}=\eta_{AB}\theta^A_\mu \theta^B_\nu$\,. 
At the same time, $f_{\mu\nu}$ can be expanded with respect to 
$\omega^A_\mu$ from \eqref{tetrad}. The 
two sets of 1-forms may 
 differ from each other by a local 
Lorentz boost, so that 
\bea                         \label{boost} 
\omega^0=\theta^0\sec\alpha+\theta^1\tan\alpha,~~
\omega^1=\theta^1\sec\alpha+\theta^0\tan\alpha,~~~~
\eea
where $\alpha$ is the boost parameter. 
Using the explicit expressions for $\omega^A$ and $\theta^A$ and 
comparing the coefficients in front of $dt,dr$ in \eqref{boost} gives four 
conditions, which determine $\alpha$ and $a$, $b$, $c$
in terms of $\Delta,T,U$.  
As a result,
the consistency condition \eqref{uuu} assumes the form 
\be                       \label{PDE}
\dot{U}T^\prime -\dot{T}U^\prime -u^2\Q^2+u\Q\sqrt{A_{+}A_{-}}/\Delta =0
\ee
with
$A_{\pm}=\Delta^2\dot{T}+U^\prime \pm (\Delta^2 T^\prime+\dot{U})$.
This equation determines $T(t,r)$. 

Let us first consider the $\eta\to 0$ limit, when $\tilde{\Lambda}=0$
and $\Delta=1$, in which case exact solutions of \eqref{PDE} can be found. 
For $k=0$, when $U=u\,\Q\, r$, we find
\be                    \label{T1}
T(t,r)=C\int^t \frac{\Q^2}{\dot{\Q}}\, dt+\left(\frac{u^2}{4C}+Cr^2 \right)\Q\,,
\ee
where $C$ is an integration constant. This solution agrees with the one
obtained in \cite{gaba} for $c_3=c_4=0$, $u=3/2$. 

For $k=1$ one has $U=u\,\Q\,\sin r$ and we find
\be                   \label{T2}
T(t,r)=\int^t \sqrt{(C^2+u^2)(\dot{\Q}^2+\Q^2)}\, dt+C\Q\cos(r).
\ee
For $k=-1$ and $U=u\,\Q\,\sinh r$ we obtain
\be                   \label{T3}
T(t,r)=\int^t \sqrt{(C^2-u^2)(\dot{\Q}^2-\Q^2)}\, dt+C\Q\cosh(r).
\ee

If $\Delta\neq 1$ then exact solutions are difficult to find,
however, at least when $\tilde{\Lambda}$ is small, the solution  can be constructed 
perturbatively as
\be
T=T_0+\sum_{n\geq 1}(-\tilde{\Lambda}/3)^n T_n. 
\ee
Here $T_0$ corresponds to the zero order expressions \eqref{T1}--\eqref{T3},
while the corrections $T_n$ can be obtained by separating the variables with 
the ansatz $T_n=\sum_{m=0}^{n+1} f_m(t) g^m(r)$ 
where $g(r)=\{r^2,\cos(r),\cosh(r)\}$ for $k=0,1,-1$, respectively. 
For example, for $k=0$ 
one finds 
\be
T_1=C\int^t \frac{\Q^6}{4\dot{\Q}^3}\, dt+\left(\frac{Cr^4}{4}+\frac{u^2r^2}{8C}
-\frac{u^4}{192 C^3}
 \right)\Q^3,
\ee
with $C$ being the same as in \eqref{T1}; similarly for $n>1$. 

This completes out constructions, since all field equations and the 
consistency condition are now fulfilled.

\noindent
{\bf Discussion--}
We have obtained the cosmological solution
in the ghost-free bigravity with matter,
for arbitrary values of the theory parameters
(provided that $u$ in \eqref{u} is real). One can choose $c_3,c_4$ 
such that $\Lambda>0$.
The metric $g_{\mu\nu}$ then describes a FRW
universe which can be spatially flat, open, or closed.
It is matter-dominated
at early times, but  at late times it enters the accelerated phase 
due to the effective cosmological term mimicked by the graviton mass.  
The metric $f_{\mu\nu}$ is the static (anti)de Sitter \eqref{AdS} parameterized by 
$T,U$. 

The two metrics are not simultaneously diagonal and do not have the same 
Killing symmetries. In particular, the translational symmetries of $g_{\mu\nu}$
are not shared by $f_{\mu\nu}$, since 
the Stueckelberg fields $\phi^0=T(t,r)$, $\phi^k=U(t,r)\{\sin\vartheta\cos\varphi,
\sin\vartheta\sin\varphi,\cos\varphi\}$ 
are inhomogeneous 
functions of $x^\mu$. One can therefore expect the fluctuations around the 
background solutions to be inhomogeneous, but 
this effect will be sourced by the terms proportional to $m^2$ in Eqs.\eqref{e1},\eqref{e2}
and so will be small in the regions smaller than $1/m$, in agreement 
with the arguments of \cite{gaba}.

Setting the matter density to zero, the solution for $g_{\mu\nu}$ is 
pure de Sitter, rewriting which in the static coordinates reproduces 
the static bigravity solutions found in \cite{Volkov}.
%for generic $c_3,c_4,\eta$. %, whose $\eta\to 0$ limit was described in \cite{Koyama}.  

The solution exists for any value of $\eta$. 
When $\eta\to 0$ then $\tilde{\Lambda}\to 0$, so that  
$f_{\mu\nu}$ becomes flat, while  
$g_{\mu\nu}$ still describes the expanding universe. 
Therefore, we obtain in this limit the solution  
of the ghost-free massive gravity with a flat reference metric \cite{RGT},
with the Stueckelberg fields expressed by \eqref{T1}--\eqref{T3}. 

Our solution therefore describes the most general 
cosmology with the FRW $g_{\mu\nu}$
but with inhomogeneous $f_{\mu\nu}$, 
and it equally applies within the 
bigravity  and massive gravity contexts.  
The latter property is quite special,
since in general the metric $f_{\mu\nu}$ does not necessary become flat 
for $\eta\to 0$, while generic 
massive gravity solutions are not guaranteed to extend to the bigravity \cite{Visser}. 

All other known cosmological solutions can be obtained by setting  
$c=0$ in eqs.\eqref{tetrad}--\eqref{LLL}, in which case both metrics are FRW.  
The corresponding bigravity solutions do not always show the late time 
acceleration and  do not cover the massive gravity case \cite{cosm1},\cite{cosm1a}.
The massive gravity cosmologies with two FRW metrics   
do not extend to the bigravity
and show a non-linear instability \cite{Muk}. 

It seems therefore that our solution is the 
most sensible physically, since it covers all possible cases
and shows the late time acceleration expected for massive gravitons.

{\bf Note added--} 
When this text was being completed, there appeared the article \cite{Wyman}
on massive gravity ($\eta=0$) cosmologies  whose analysis is 
partly similar to the 
above discussion, although it 
does not give the explicit solution for the Stueckelberg scalars.

\end{document}